\documentclass[preprint2]{aastex}
\usepackage{graphicx}

\shorttitle{Electron-Impact Excitation of S\,{\sc iii}}
\shortauthors{Ramsbottom, Grieve \& Hudson}

\begin{document}

\title{Electron-Impact Excitation Collision Strengths and Theoretical Line Intensities for Fine-Structure Transitions in S\,{\sc iii}}
\author{M. F. R. Grieve, C. A. Ramsbottom and C. E. Hudson}
\affil{Centre for Theoretical Atomic, Molecular and Optical Physics, School of Mathematics and Physics, Queen's University Belfast,\\
    Belfast, BT7 1NN, UK}
    \email{c.ramsbottom@qub.ac.uk}
        \author{F.P. Keenan}
    \affil{Astrophysics Research Centre, School of Mathematics and Physics, Queen's University Belfast, Belfast, BT7 1NN, UK}

\begin{abstract}
We present Maxwellian-averaged effective collision strengths for the electron-impact excitation of S\,{\sc iii} over 
a wide range of electron temperatures of astrophysical importance, log T$_{e}$(K) = 3.0--6.0. The calculation incorporates 53 fine-structure
levels arising from the six configurations 3s$^2$3p$^2$, 3s3p$^3$, 3s$^2$3p3d, 3s$^2$3p4s, 3s$^2$3p4p and 3s$^2$3p4d, giving rise to 1378 individual lines, and is undertaken 
using the recently developed RMATRX II plus FINE95 suite of codes. 
A detailed comparison is made with a previous $R$-matrix calculation and significant differences are found 
for some transitions. 
The atomic data are subsequently incorporated into the modeling code {\sc cloudy} to generate line intensities for a range of plasma parameters, with emphasis on allowed UV and EUV emission lines detected from the Io plasma torus. Electron density-sensitive line ratios are calculated with the present atomic data 
and compared with those from CHIANTI v7.1, as well as with Io plasma torus spectra 
obtained by FUSE and EUVE. The present line intensities are found to agree well with the observational results and provide a noticeable improvement upon the values predicted by CHIANTI.
\end{abstract}



\section{Introduction}

Emission lines arising from transitions in the Si-like ion S\,{\sc iii} have been widely detected in the IR, optical and UV spectral regions of astronomical objects (see, for example, Simpson et al. 1997; Binette et al. 2012; Feldman et al. 2004). This has led to a number of calculations over the years of collisional atomic data for this ion, which can be used to determine theoretical emission line intensity ratios to infer electron temperature, density and chemical abundance in a host of plasma sources. Two $R$-matrix evaluations were performed in the 1990's, namely by 
Galav\'{i}s et al. (1995) who incorporated the lowest 17 levels corresponding to the 3s$^2$3p$^2$, 3s3p$^3$ and 3s$^2$3p3d levels, while Tayal \& Gupta (1999) extended this to 27 LS states by including the 3s$^2$3p4s, 3s$^2$3p4p and 3s$^2$3p4d configurations. Both of these calculations were performed in LS coupling, with the fine-structure data generated by transforming the LS coupled $K$-matrices to an intermediate coupling scheme. Unfortunately, for all three of the important IR transitions among the 3s$^2$3p$^2$ $^3$P$_J$ fine-structure levels of the ground state configuration, the effective collision strengths calculated by both groups of authors 
differed significantly in behaviour and magnitude. It was not until recently 
that these differences were resolved by Hudson et al. (2012), who performed a further $R$-matrix calculation  incorporating 53 fine-structure target levels with configurations 3s$^2$3p$^2$, 3s3p$^3$, 3s$^2$3p3d, 3s$^2$3p4s, 3s$^2$3p4p and 3s$^2$3p4d. The new RMATRX II plus FINE95 (Burke et al. 1994, Burke 2012) suite of programs were adopted in this work, and the effective collision strengths produced agreed with the earlier but less sophisticated IRON Project data from Galav\'{i}s et al. (1995), while the source of the discrepancy with Tayal \& Gupta was also discussed. Only the IR transitions within the ground configuration of S\,{\sc iii} were presented and analysed in these 3 papers, 
all of which represent low-lying forbidden lines. The question thus arises as to the reliability of the Tayal \& Gupta data for transitions among levels higher than the ground state, and particularly for the low-lying allowed lines of astrophysical importance. None of these lines were considered in the Galav\'{i}s et al. (1995) paper.

The CHIANTI database (Dere et al. 1997, Landi et al. 2012) is a primary source of atomic data employed by astrophysicists and plasma physicists for the evaluation of temperature and density diagnostic line ratios and other modeling applications. Currently, CHIANTI (Version 7.1) stores the effective collision strengths of Tayal \& Gupta (1999) as spline fits scaled according to the rules formulated by Burgess \& Tully (1992). It is these data which are incorporated into astrophysical models at present. For example, Feldman et al. (2004) obtained 
the spectrum of the Io plasma torus in the wavelength range 905--1187\,\AA\ using the Far Ultraviolet Spectroscopic Explorer (FUSE) and found this to be dominated by transitions of S\,{\sc ii}, S\,{\sc iii}
 and S\,{\sc iv}, many of which are extremely useful as temperature and/or density diagnostics. In particular, they found that two density sensitive line ratios (S\,{\sc iii} 1021/1012 and S\,{\sc iii} 1077/1012) gave very 
 different values for n$_{e}$ due to the fact that the calculated ratios were extremely sensitive to small uncertainties in the CHIANTI atomic data parameters. Additionally, Dong \& Draine (2011) adopted the collisional rate coefficients of Tayal \& Gupta (1999) in their model to investigate the evolution of temperature, ionization and emission from cooling in the warm ionized medium (WIM). A total of 159 lines were included in the model, 26 of which accounted for over 95$\%$ of the radiative cooling. Of these important lines, 4 were of S\,{\sc iii} (namely 33.48$\mu$m, 18.71$\mu$m, 9533\,\AA\ and 9071\,\AA). Finally, Otsuka et al. (2010) examined the electron temperature and density structure within the halo planetary nebula BoBn1 using 16 diagnostic line ratios,  with the electron temperature estimated in the [S\,{\sc iii}] zone for the first time. Again, atomic data for the S\,{\sc iii} ion were taken from Tayal \& Gupta (1999).
 
It is hence important that the atomic data currently stored in CHIANTI is thoroughly tested against the new results
of Hudson et al. (2012), particularly for transitions to the higher-lying levels of significant astrophysical importance. The S\,{\sc iii} ion is well known as one of the most prominent emitters in the Io plasma torus, which arises due to volcanic activity on Io. A large number of S\,{\sc iii} transitions have been detected in the EUV wavelength region by several satellite missions. Hall et al. (1994b) analysed sulfur ion emission from the Io plasma observed by the Hopkins Ultraviolet Telescope (HUT) at low spectral resolution over the 
wavelength regions 830--1864\,\AA\ in first order and 415--932\,\AA\ in second. Subsequently, Hall et al. (1994b) presented Extreme Ultraviolet Explorer (EUVE) satellite observations of the Jupiter system which showed a rich emission line spectrum from the Io plasma torus in the 370--735\,\AA\ wavelength region. Finally, Feldman et al. (2004) recorded the spectrum of the Io plasma torus between 905--1187\,\AA\ using FUSE. In Table 1 we list all of the identified 
S\,{\sc iii} emission lines observed by HUT, EUVE and FUSE. It can be seen from the table that many of these
are allowed transitions from the ground and metastable levels to the higher-lying 4s, 4p and 4d states. Unfortunately,
none of these lines were considered by Hudson et al. (2012), and no attempt was made to assess 
the accuracy of the atomic data currently available 
and how they compare with the Tayal \& Gupta (1999) results currently in CHIANTI. 

This paper is constructed as follows. In Section 2 the details 
of the collision calculation are provided, while in Section 3 we present a graphical analysis of the results for a number of significant transitions and compare these with previous theoretical work. The convergence of the collision strengths for the allowed lines is also investigated. Finally, in Section 4 we use CLOUDY to produce theoretical line intensity ratios for the Io plasma torus, and compare these 
with existing observational data from FUSE and EUVE spectra. The current theoretical line ratios are also compared with 
values generated from CHIANTI Version 7.1, and an assessment made as to which atomic dataset produces the most reliable results.

\begin{deluxetable}{cccccc}
\tablecaption{EUV emission lines of S\,{\sc iii} identified in HUT (Hall et al. 1994b), EUVE (Hall et al. 1994a) and FUSE (Feldman et al. 2004) spectra of the Io torus.\label{table:emmisionfeatures}}
\tablewidth{0pt}
\tablehead{
\colhead{Mission} && \colhead{Wavelength \,(\AA)} && \colhead{Transition} 
}
\startdata
HUT   &&  1713, 1729   && 3s$^2$3p$^2\;^3$P$^e$ - 3s3p$^3\;^5$S$^{\rm{o}}$ \\
      &&  1190 - 1202  && 3s$^2$3p$^2\;^3$P$^e$ - 3s3p$^3\;^3$D$^{\rm{o}}$ \\
      && 1012 - 1022   && 3s$^2$3p$^2\;^3$P$^e$ - 3s3p$^3\;^3$P$^{\rm{o}}$ \\
      &&  1077         && 3s$^2$3p$^2\;^1$D$^e$ - 3s$^2$3p3d $^1$D$^{\rm{o}}$ \\
      &&  724 - 729    && 3s$^2$3p$^2\;^3$P$^e$ - 3s3p$^3\;^3$S$^{\rm{o}}$ \\
      &&  700 - 703    && 3s$^2$3p$^2\;^3$P$^e$ - 3s$^2$3p3d $^3$P$^{\rm{o}}$ \\
   \\ \hline
   \\
   
EUVE  &&  480 - 489    && 3s$^2$3p$^2\;^3$P$^e$ - 3s$^2$3p4d$^3$P$^{\rm{o}}$ \\
      &&               && 3s$^2$3p$^2\;^3$P$^e$ - 3s$^2$3p4d$^3$D$^{\rm{o}}$ \\
      &&  671 - 685    && 3s$^2$3p$^2\;^3$P$^e$ - 3s$^2$3p3d$^3$D$^{\rm{o}}$ \\
      &&               && 3s$^2$3p$^2\;^3$P$^e$ - 3s$^2$3p4s$^3$P$^{\rm{o}}$ \\
      &&  679 - 705    && 3s$^2$3p$^2\;^3$P$^e$ - 3s$^2$3p3d$^3$P$^{\rm{o}}$ \\
      &&  724 - 731    && 3s$^2$3p$^2\;^3$P$^e$ - 3s3p$^3\;^3$S$^{\rm{o}} $\\
      &&               && 3s$^2$3p$^2\;^1$D$^e$ - 3s$^2$3p4s$^1$P$^{\rm{o}}$ \\
   \\ \hline
   \\
FUSE  &&  1012.492     && 3s$^2$3p$^2\;^3$P$^e_{0}$ - 3s3p$^3\;^3$P$^{\rm{o}}_{1}$ \\
      &&  1015.496     && 3s$^2$3p$^2\;^3$P$^e_{1}$ - 3s3p$^3\;^3$P$^{\rm{o}}_{0}$ \\
      &&  1015.561     && 3s$^2$3p$^2\;^3$P$^e_{1}$ - 3s3p$^3\;^3$P$^{\rm{o}}_{1}$ \\
      &&  1015.775     && 3s$^2$3p$^2\;^3$P$^e_{1}$ - 3s3p$^3\;^3$P$^{\rm{o}}_{2}$ \\
      &&  1021.105     && 3s$^2$3p$^2\;^3$P$^e_{2}$ - 3s3p$^3\;^3$P$^{\rm{o}}_{1}$ \\
      &&  1021.321     && 3s$^2$3p$^2\;^3$P$^e_{2}$ - 3s3p$^3\;^3$P$^{\rm{o}}_{2}$ \\
      &&  1077.171     && 3s$^2$3p$^2\;^1$D$^e_{2}$ - 3s$^2$3p3d$^1$D$^{\rm{o}}_{2}$ \\
      &&  1121.755     && 3s3p$^3\;^3$D$^{\rm{o}}_{2}$ - 3s$^2$3p4p$^3$P$^e_{2}$ \\
      &&  1122.418     && 3s3p$^3\;^3$D$^{\rm{o}}_{3}$ - 3s$^2$3p4p$^3$P$^e_{2}$ \\
      &&  1126.536     && 3s3p$^3\;^3$D$^{\rm{o}}_{1}$ - 3s$^2$3p4p$^3$P$^e_{1}$ \\
      &&  1126.886     && 3s3p$^3\;^3$D$^{\rm{o}}_{2}$ - 3s$^2$3p4p$^3$P$^e_{1}$ \\
      &&  1128.500     && 3s3p$^3\;^3$D$^{\rm{o}}_{1}$ - 3s$^2$3p4p$^3$P$^e_{0}$ \\
\enddata
\end{deluxetable}

\section{The target model}

The target model used in the present calculation has been described in detail by Hudson et al. (2012) and will only be summarized here for clarity and completeness. A total of 29 $LS$ states (or 53 fine-structure levels), formed from the basis configurations 3s$^2$3p$^2$, 3s3p$^3$, 3s$^2$3p3d, 3s$^2$3p4s, 3s$^2$3p4p and 3s$^2$3p4d, were included in the wavefunction representation of the S\,{\sc iii} target model. All singlet, triplet and quintet spin states were included and represented by configuration-interaction expansions in terms of nine orthogonal basis orbitals, eight spectroscopic (1s, 2s, 2p, 3s, 3p, 3d, 4s, 4p) and one pseudo-oribital ($\overline{4{\rm d}}$) which allowed for additional correlation effects. Outside of the Hartree-Fock sea the parameters for all orbitals were generated using the configuration-interaction CIV3 code of Hibbert (1975). The accuracy of this S\,{\sc iii} target model has been thoroughly analysed by Hudson et al. (2012), where it was found that the even parity levels differed from the observed NIST values by at most 5$\%$ (with the exception of the 3s$^2$3p$^2\;^1$D$^e$ level), while for the odd parity levels the discrepancies were less than 1$\%$ for the majority of levels, apart from 3s3p$^3\;^5$S$^{\rm{o}}$ 
where a difference of 11$\%$ was noted. Transitions to these odd parity states from the ground and excited levels of the 3s$^2$3p$^2$ configuration are the focus of the present investigation. In Table 2 the 53 fine-structure levels are 
listed, with each assigned an index which will subsequently be used when 
denoting a particular transition between two levels.

\begin{deluxetable}{cc|cc}
\tablecaption{Index of the S\,{\sc iii} fine-structure levels.\label{table:indexvalues}}
\tablewidth{0pt}
\tablehead{
\colhead{Index} & \colhead{Level} & \colhead{Index} & \colhead{Level}
}
\startdata
1 & 3s$^2$3p$^2\,^3$P$^{\rm e}_0$   &  28 & 3s$^2$3p4s\,$^1$P$^{\rm o}_1$ \\
2 & 3s$^2$3p$^2\,^3$P$^{\rm e}_1$   &  29 & 3s3p$^3\,^1$D$^{\rm o}_2$  \\
3 & 3s$^2$3p$^2\,^3$P$^{\rm e}_2$   &  30 & 3s$^2$3p3d\,$^1$F$^{\rm o}_3$ \\
4 & 3s$^2$3p$^2\,^1$D$^{\rm e}_2$   &  31 & 3s$^2$3p3d\,$^1$P$^{\rm o}_1$ \\
5 & 3s$^2$3p$^2\,^1$S$^{\rm e}_0$   &  32 & 3s$^2$3p4p\,$^1$P$^{\rm e}_1$ \\
6 & 3s3p$^3\,^5$S$^{\rm o}_2$       &  33 & 3s$^2$3p4p\,$^3$D$^{\rm o}_1$ \\
7 & 3s3p$^3\,^3$D$^{\rm o}_1$       &  34 & 3s$^2$3p4p\,$^3$D$^{\rm o}_2$  \\
8 & 3s3p$^3\,^3$D$^{\rm o}_2$       &  35 & 3s$^2$3p4p\,$^3$D$^{\rm o}_3$ \\
9 & 3s3p$^3\,^3$D$^{\rm o}_3$       &  36 & 3s$^2$3p4p\,$^3$P$^{\rm e}_0$ \\
10 & 3s3p$^3\,^3$P$^{\rm o}_2$      &  37 & 3s$^2$3p4p\,$^3$P$^{\rm e}_1$ \\
11 & 3s3p$^3\,^3$P$^{\rm o}_1$      &  38 & 3s$^2$3p4p\,$^3$P$^{\rm e}_2$ \\
12 & 3s3p$^3\,^3$P$^{\rm o}_0$      &  39 & 3s$^2$3p4p\,$^3$S$^{\rm e}_1$ \\
13 & 3s$^2$3p3d\,$^1$D$^{\rm o}_2$  &  40 & 3s$^2$3p4p\,$^1$D$^{\rm e}_2$ \\
14 & 3s$^2$3p3d\,$^3$F$^{\rm o}_2$  &  41 & 3s$^2$3p4p\,$^1$S$^{\rm e}_0$ \\
15 & 3s$^2$3p3d\,$^3$F$^{\rm o}_3$  &  42 & 3s$^2$3p4d\,$^3$F$^{\rm o}_2$ \\
16 & 3s$^2$3p3d\,$^3$F$^{\rm o}_4$  &  43 & 3s$^2$3p4d\,$^3$F$^{\rm o}_3$  \\
17 & 3s3p$^3\,^1$P$^{\rm o}_1$      &  44 & 3s$^2$3p4d\,$^1$D$^{\rm o}_2$ \\
18 & 3s3p$^3\,^3$S$^{\rm o}_1$      &  45 & 3s$^2$3p4d\,$^3$F$^{\rm o}_4$ \\
19 & 3s$^2$3p3d\,$^3$P$^{\rm o}_0$  &  46 & 3s$^2$3p4d\,$^3$D$^{\rm o}_1$  \\
20 & 3s$^2$3p3d\,$^3$P$^{\rm o}_1$  &  47 & 3s$^2$3p4p\,$^3$D$^{\rm o}_2$ \\
21 & 3s$^2$3p3d\,$^3$P$^{\rm o}_2$  &  48 & 3s$^2$3p4p\,$^3$D$^{\rm o}_3$ \\
22 & 3s$^2$3p4s\,$^3$P$^{\rm o}_0$  &  49 & 3s$^2$3p4d\,$^3$P$^{\rm o}_2$  \\
23 & 3s$^2$3p4s\,$^3$P$^{\rm o}_1$  &  50 & 3s$^2$3p4d\,$^3$P$^{\rm o}_1$  \\
24 & 3s$^2$3p3d\,$^3$P$^{\rm o}_2$  &  51 & 3s$^2$3p4d\,$^3$P$^{\rm o}_0$ \\
25 & 3s$^2$3p3d\,$^3$D$^{\rm o}_1$  &  52 & 3s$^2$3p4d\,$^1$F$^{\rm o}_3$  \\
26 & 3s$^2$3p3d\,$^3$D$^{\rm o}_2$  &  53  & 3s$^2$3p4d\,$^1$P$^{\rm o}_1$ \\
27 & 3s$^2$3p3d\,$^3$D$^{\rm o}_3$  &  \\
\enddata
\end{deluxetable}

Collision calculations were initially undertaken in $LS$ coupling using the parallel RMATRX II $R$-matrix packages (Burke et al. 1994) in the internal region. The resulting diagonalized Hamiltonian matrices were then transformed to J$\pi$ coupling, via the transformation code FINE95 (Burke, private communication), on the external boundary by employing the term-coupling coefficients to take account of the term splitting in the target. Finally, the parallel Breit-Pauli codes developed by Ballance \& Griffin (2004) were used to compute the electron impact excitation cross sections at a very fine mesh of incident electron energies in the external region ($\approx$ 10$^{-4}$ Ryd). In the work of Hudson et al. (2012), convergence of the low-lying forbidden transitions among the 3s$^2$3p$^2$ ground state configuration was achieved by including (N+1)-electron symmetries up to 2J = 9, with contributions from higher partial waves being negligible. Due to the long range nature of the Coulomb potential, contributions from higher partial waves are found to be significant for the optically allowed transitions of interest in this paper. Hence, additional partial waves up to 2J = 25 for all doublet, quartet and sextet spin states with even and odd parities were incorporated into the present calculation to allow for these further contributions to the dipole allowed transitions from the higher partial waves. In addition, a further `top-up' procedure using the Burgess sum rule for the dipole transitions and a geometric series for the non-dipole transitions was included to ensure complete convergence of the collision strengths for all allowed lines for the range of impact energies considered. The effect of this top-up and the convergence of the collision strengths will be discussed in Section 3.

For many astrophysical and plasma applications it is not the electron impact excitation collision strengths that are required but the Maxwellian-averaged effective collision strengths between the fine-structure levels. A thermally or Maxwellian-averaged effective collision strength ($\Upsilon_{if}$) between an initial state $i$ and a final state $f$ may be derived from the collision strength ($\Omega_{if}$) using the expression

\begin{equation}\label{eq:maxwell}
\Upsilon_{if}(T_e) = \int_0^{\infty}\Omega_{if}(E_f)\exp(-E_f/kT_e)d(E_f/kT_e).
\end{equation}
where $E_f$ is the final kinetic energy of the ejected electron, $T_e$ the electron temperature in Kelvin and $k$ is Boltzmann's constant.

\section{Results and discussion}

We consider first the spin-forbidden 3s$^2$3p$^2\;^3$P$^e_J$ -- 3s3p$^3\;^5$S$^{\rm{o}}_2$ intercombination multiplet observed by HUT at 1728.9\,\AA\ (J = 2-2 line) and 1713.1\,\AA\ (J= 1-2). These transitions have  
been used to calculate ion partitioning and their intensity serves as one of several temperature diagnostics for S\,{\sc iii}. We present in Figure 1 the effective collision strength as a function of 
electron temperature for the three fine-structure transitions that contribute to this spin-forbidden multiplet, namely
3s$^2$3p$^2\;^3$P$^e_{0,1,2}$ -- 3s3p$^3\;^5$S$^{\rm{o}}_2$ (1-6, 2-6, 3-6). Also shown are the calculations of 
Tayal \& Gupta (1999) for a range of temperatures spanning that of maximum abundance in ionisation equilibrium for the S\,{\sc iii} ion, log T$_e$ = 4.8. 
Agreement is quite good at many temperatures, but for the lower values 
the two calculations deviate significantly. For example, at the lowest temperature considered by Tayal \& Gupta (1999), log T$_e$ = 3.7, the discrepancies range from 40--50$\%$.

\begin{figure}[h]
\centering
\includegraphics[width=\hsize]{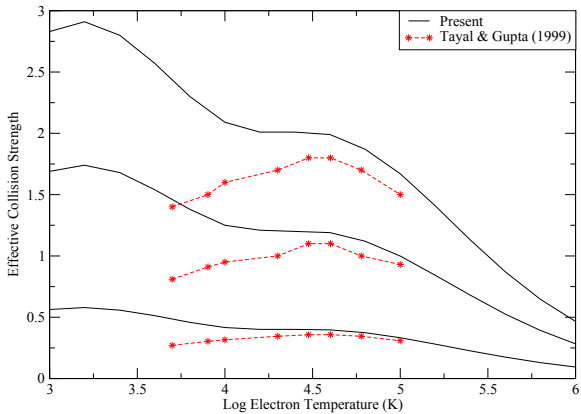}
\caption{Effective collision strength for transitions 1, 2, 3 -- 6 (3s$^2$3p$^2\,^3$P$^e_{0,1,2}$ -- 3s3p$^3\;^5$S$^{\rm{o}}_2$). The present results are shown by a solid black line while the red stars represent the data of
Tayal \& Gupta (1999).}
\label{fig1} 
\end{figure}

In Figure 2 we plot similar results for the 
3s$^2$3p$^2\;^3$P$^e_{0,1,2}$ -- 3s3p$^3\;^3$S$^{\rm{o}}_1$ (1-18, 2-18, 3-18) allowed fine-structure transitions observed by both HUT and EUVE in the 724 -- 729\,\AA\ wavelength range. There is good agreement between the present calculations and those of Tayal \&\ Gupta (1999). Also shown in the figure are the calculations of Hudson et al. (2012), who only included the low partial wave contributions up to 2J = 9 to allow for convergence of the low-lying forbidden lines.
It is clear from the figure that the higher partial waves included in the present work, as well as the `top-up' from even higher partial waves, are significant and must be included to ensure convergence of these slowly-converging allowed transitions. At the highest temperature considered (log T$_e$ = 6.0) the higher partial wave contributions enhance the effective collision strengths by a factor of 2.

\begin{figure}[h]
\centering
\includegraphics[width=\hsize]{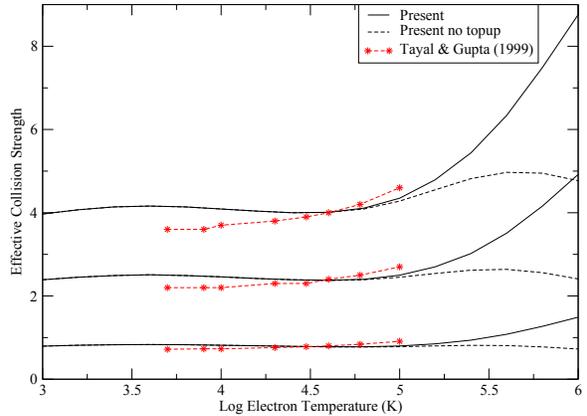}
\caption{Same as Figure 1 except for transitions 1, 2, 3 -- 18 (3s$^2$3p$^2\,^3$P$^e_{0,1,2}$ -- 3s3p$^3\;^3$S$^{\rm{o}}_1$). However, also shown in the figure are the present results with no top-up from higher partial waves.}
\label{fig2} 
\end{figure}

In Figure 3 we present results for the dipole-allowed 3s$^2$3p$^2\;^3$P$^e_1$ -- 3s$^2$3p4s $^3$P$^{\rm{o}}_{0,1,2}$ (2-22, 2-23, 2-24) transitions. From Table 1 we see that these lines were observed by EUVE in the 671 -- 685\,\AA\ range (Hall et al. 1994b). Significant differences are evident between the present results and those of Tayal \& Gupta (1999) for all three lines shown in the figure. Discrepancies of up to 50$\%$ are found, with the Tayal \&\ Gupta (1999)
data being significantly smaller than the present results at all temperatures where a comparison is possible. The implications of these differing atomic data on the resultant diagnostics  will be discussed in detail in Section 4. 

\begin{figure}[h]
\centering
\includegraphics[width=\hsize]{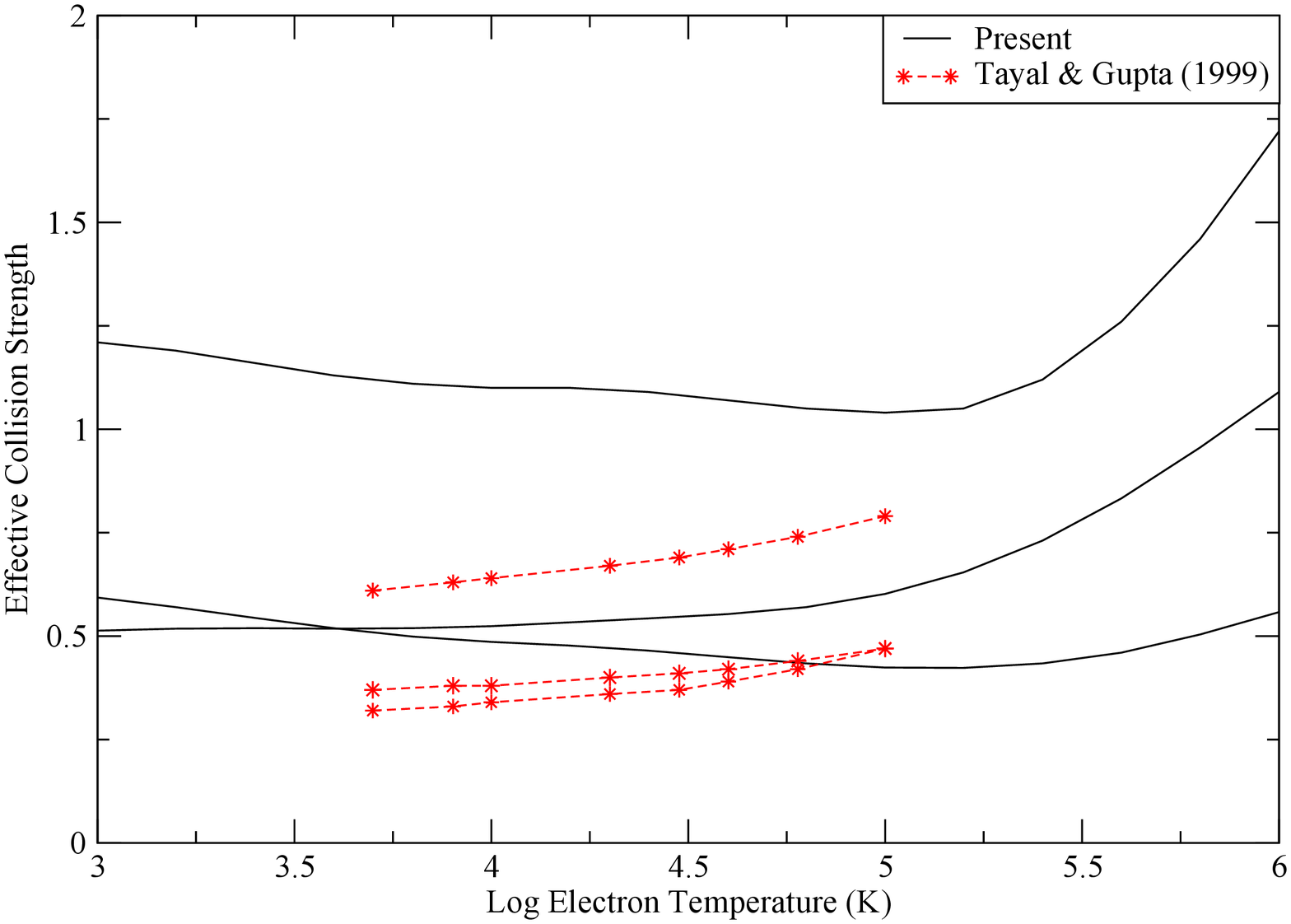}
\caption{Same as Figure 1 except for transitions 2 -- 22, 23, 24 (3s$^2$3p$^2\,^3$P$^e_1$ -- 3s$^2$3p4s $^3$P$^{\rm{o}}_{0,1,2}$).}
\label{fig3} 
\end{figure}

We now turn our attention to the 1012--1022\,\AA\ wavelength range in FUSE observations, where several 
S\,{\sc iii} transitions have been identified (Feldman et al. 2004). In Figure 4 we present
effective collision strengths as a function of temperature for transitions 3s$^2$3p$^2\;^3$P$^e_2$ -- 3s3p$^3\;^3$P$^{\rm{o}}_{0,1,2}$ (3-10, 3-11, 3-12). Excellent agreement is found between the present results and those of Tayal \&\ Gupta (1999) at all temperatures considered, with discrepancies of only a few per cent for all 3 transitions. 

\begin{figure}[h]
\centering
\includegraphics[width=\hsize]{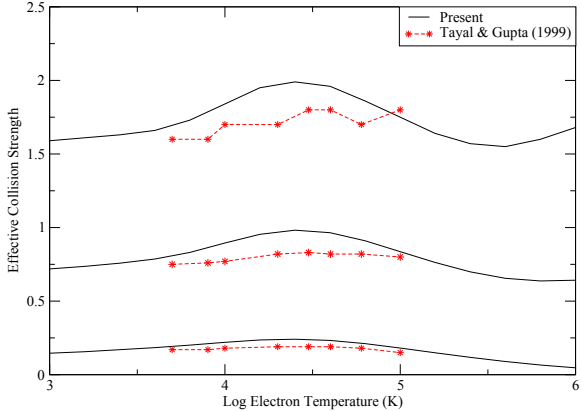}
\caption{Same as Figure 1 except for transitions 3 -- 10, 11, 12 (3s$^2$3p$^2\,^3$P$^e_2$ -- 3s3p$^3 \; ^3$P$^{\rm{o}}_{0,1,2}$).}
\label{fig4} 
\end{figure}

Finally, we investigate some of the higher-lying (3s$^2$3p4d) levels. Transitions from the ground state to these
have been observed by EUVE in the 480 -- 489\,\AA\ wavelength range. We present in Figure 5 
effective collision strengths for the 3s$^2$3p$^2\;^3$P$^e_2$ -- 3s$^2$3p4d $^3$P$^{\rm{o}}_{0,1,2}$ (3-49, 3-50, 3-51) transitions. Significant differences are found with the Tayal \& Gupta (1999) data at 
all temperatures considered, up to a factor of 5 for the strongest J = 2-2 line. We note that there are 
similar discrepancies for other transitions involving the high-lying 3s$^2$3p4d levels. These differing atomic datasets will affect the theoretical diagnostics, as discussed in \S\ 4.

\begin{figure}[h]
\centering
\includegraphics[width=\hsize, trim = 10mm 0mm 0mm 0mm ]{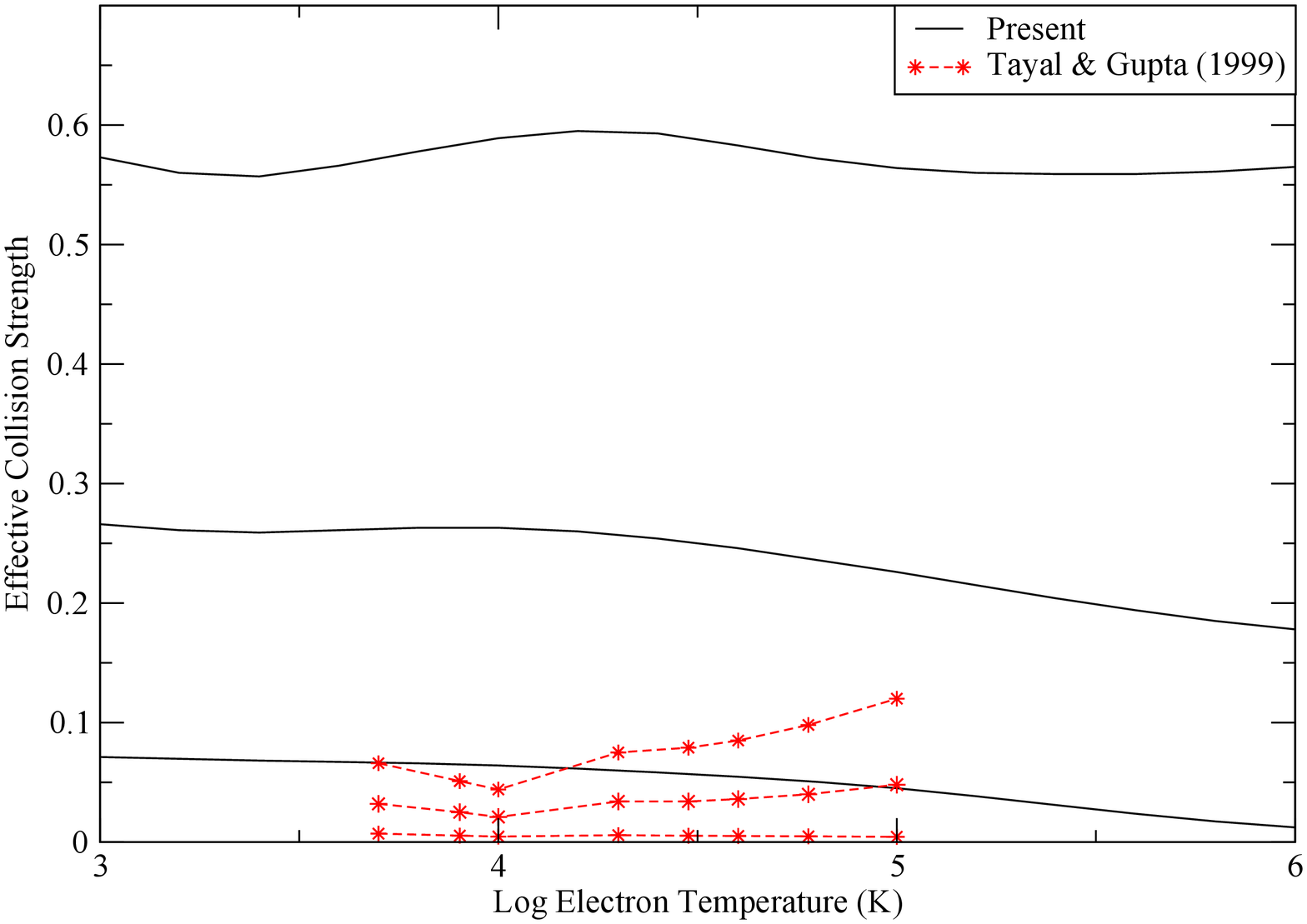}
\caption{Same as Figure 1 except for transitions 3 -- 49, 50, 51 (3s$^2$3p$^2\,^3$P$^e_2$ -- 3s$^2$3p4d $^3$P$^{\rm{o}}_{0,1,2}$).}
\label{fig5} 
\end{figure}

In summary, the level of agreement between the present results and the earlier work of Tayal \& Gupta (1999) is varied. For several of the transitions the agreement is excellent (highlighted in Figures 2 and 4), but for many of the astrophysically important lines the discrepancies are significant (highlighted in Figures 1, 3 and 5). We present in Table \ref{ECS} the complete set of effective collision strengths for all 1378 transitions 
among the 53 fine-structure levels considered in this work, at 16 temperatures ranging from log T$_e$ = 3.0--6.0. Note that this table is only available on-line, with the hardcopy version containing a portion of the results for illustrative purposes.

\begin{deluxetable}{c c c c c c c c c c}
\tablewidth{0 pt}
\tablecaption{Effective collision strengths for S\,{\sc iii} as a function of temperature \label{ECS}}
\tablehead{
$i$ & $j$ &  \multicolumn{8}{c}{Temperature (K)}  \\
\hline
  &      & $ 1.00^{+03}$ & $ 1.58^{+03}$ & $ 2.51^{+03}$ & $ 3.98^{+03}$ & $ 6.31^{+03}$ & $ 1.00^{+04}$ & $ 1.58^{+04}$ & $ 2.51^{+04}$\\
     &      & $ 3.98^{+04}$ & $ 6.31^{+04}$ & $ 1.00^{+05}$ & $ 1.58^{+05}$ & $ 2.51^{+05}$ & $ 3.98^{+05}$ & $ 6.31^{+05}$ & $ 1.00^{+06}$}
\startdata
   1 &    2 & $ 2.08^{+00}$ & $ 2.10^{+00}$ & $ 2.13^{+00}$ & $ 2.20^{+00}$ & $ 2.27^{+00}$ & $ 2.26^{+00}$ & $ 2.17^{+00}$ & $ 2.07^{+00}$\\
     &      & $ 1.97^{+00}$ & $ 1.84^{+00}$ & $ 1.63^{+00}$ & $ 1.36^{+00}$ & $ 1.07^{+00}$ & $ 8.10^{-01}$ & $ 5.90^{-01}$ & $ 4.18^{-01}$\\
   1 &    3 & $ 9.84^{-01}$ & $ 9.59^{-01}$ & $ 9.47^{-01}$ & $ 9.61^{-01}$ & $ 9.85^{-01}$ & $ 1.03^{+00}$ & $ 1.12^{+00}$ & $ 1.24^{+00}$\\
     &      & $ 1.32^{+00}$ & $ 1.32^{+00}$ & $ 1.23^{+00}$ & $ 1.07^{+00}$ & $ 8.87^{-01}$ & $ 7.10^{-01}$ & $ 5.62^{-01}$ & $ 4.45^{-01}$\\
   1 &    4 & $ 6.98^{-01}$ & $ 7.33^{-01}$ & $ 7.38^{-01}$ & $ 7.20^{-01}$ & $ 7.10^{-01}$ & $ 7.29^{-01}$ & $ 7.65^{-01}$ & $ 7.90^{-01}$\\
     &      & $ 7.86^{-01}$ & $ 7.43^{-01}$ & $ 6.59^{-01}$ & $ 5.48^{-01}$ & $ 4.30^{-01}$ & $ 3.22^{-01}$ & $ 2.34^{-01}$ & $ 1.65^{-01}$\\
   1 &    5 & $ 8.39^{-02}$ & $ 8.98^{-02}$ & $ 9.64^{-02}$ & $ 1.04^{-01}$ & $ 1.14^{-01}$ & $ 1.25^{-01}$ & $ 1.39^{-01}$ & $ 1.56^{-01}$\\
     &      & $ 1.73^{-01}$ & $ 1.82^{-01}$ & $ 1.76^{-01}$ & $ 1.60^{-01}$ & $ 1.37^{-01}$ & $ 1.14^{-01}$ & $ 9.23^{-02}$ & $ 7.37^{-02}$\\
        1 &    6 & $ 5.63^{-01}$ & $ 5.79^{-01}$ & $ 5.58^{-01}$ & $ 5.13^{-01}$ & $ 4.58^{-01}$ & $ 4.16^{-01}$ & $ 4.01^{-01}$ & $ 4.01^{-01}$\\
     &      & $ 3.97^{-01}$ & $ 3.74^{-01}$ & $ 3.33^{-01}$ & $ 2.81^{-01}$ & $ 2.26^{-01}$ & $ 1.75^{-01}$ & $ 1.30^{-01}$ & $ 9.38^{-02}$\\
       \enddata
\tablecomments{Table \ref{ECS} is published online in its entirety in the electronic edition of the Astrophysical Journal. Only a portion is shown here for illustrative purposes.}
  \end{deluxetable}


\section{Line intensity ratios}

In this section the new collision strength data are  employed, along with existing A-values, to generate 
a set of theoretical line intensity ratios for S\,{\sc iii}. Emission lines of [S\,{\sc iii}] provide well-known density and temperature diagnostics via their intensity ratios. However, it is not only the forbidden lines which are of importance, as allowed lines have also been shown to act as crucial electron density indicators.  One of the most prominent sources of S\,{\sc iii} emission is the Io plasma torus, while SUMER satellite spectra of the quiet Sun  reveal the presence of many S\,{\sc iii} lines (Curdt et al. 2004). A large proportion of nebulae, for example M\,42 and NGC\,7009 (Fang \& Liu 2011),  also show several [S\,{\sc iii}] emission features. However, in this section we will focus on S\,{\sc iii} lines observed in UV and EUV spectra of the Io plasma torus, due to the very large number of features detected, as indicated in Table 1.

The present set of collision strengths was employed, along with the A-value data from CHIANTI Version 7.1 (Froese Fischer et al. 2006; Tayal 1997), to generate
new line ratios (denoted L-Present) using {\sc cloudy} C13 (Ferland et al. 2013).
We have also calculated ratios (denoted L-Tayal) using {\sc cloudy} C13 with all atomic data from  
Version 7.1 of the CHIANTI database, which employs the effective collision strengths of Tayal \& Gupta (1999). A
final set of line ratios (denoted L-Feldman) were derived 
using {\sc cloudy} C13 with Version 4.0 of the CHIANTI database, to reproduce the results of Feldman et al. (2004). This version of CHIANTI employs the A-values of Tayal (1997) and Huang (1985), plus the effective collision strengths of Tayal \&\ Gupta (1999). Unless indicated otherwise, an electron temperature for the Io torus of 50,000\,K was adopted (Feldman et al. 2004) in the line ratio calculations.

\begin{figure}[h]
\centering
\includegraphics[width=\hsize]{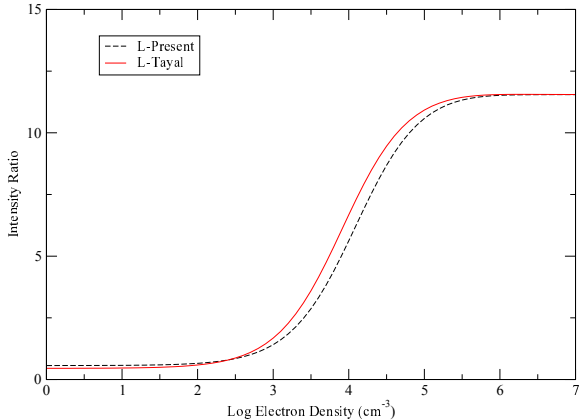}
\caption{The S {\sc iii} 18.7\,$\mu$m/33.5\,$\mu$m line intensity ratio plotted 
as a function of logarithmic electron density at a temperature of 10,000\,K. The dashed black curve indicates the 
L-Present calculations and the solid red line the L-Tayal values.}
\label{forden}
\end{figure}

Before proceeding to an examination of the allowed S\,{\sc iii} emission lines in the Io plasma torus, it is worthwhile
to look at the important forbidden transitions. Hudson et al. (2012) highlighted some noticeable differences between their effective collision strengths and the results of Tayal \& Gupta (1999) for some of the low-lying forbidden transitions. The impact these differences have on the resulting line intensity ratios will now be examined. For example, the [S {\sc iii}] 18.7\,$\mu$m and 33.5\,$\mu$m lines form an extremely important diagnostic ratio which is ideal for the analysis of low density astronomical objects.  These two lines are observed in a wide range of nebulae and are frequently employed to determine the electron density. Their intensity ratio is presented in Figure \ref{forden} at a temperature of 10,000\,K. It can be seen that there are some differences between the L-Present and L-Tayal
curves, especially for electron densities around 10$^{4}$\,cm$^{-3}$. For example, for an intensity ratio of 5.0, L-Tayal predicts an electron density of 10$^{3.7}$\,cm$^{-3}$, while L-Present indicates 10$^{3.9}$\,cm$^{-3}$. We note that at the higher temperature of T$_{e}$ = 50,000\,K, the L-Tayal and L-Present results are in good agreement. Overall, we find very good agreement between the current and previous line ratio calculations, and this comparison helps to confirm the reliability of the theoretical results for this important density diagnostic.  

\subsection{EUV emission from the Io plasma torus}

The Io plasma torus was first identified 
by the Ultraviolet Spectrographs (UVS) on board  Voyager 1 and Voyager 2 (Broadfoot et al. 1979).  Many subsequent observations have been made, revealing a wealth of emission features in the UV and EUV spectral regions arising from sulphur ions. The large abundance of S$^{2+}$ in the Io plasma torus is a result of the high levels of volcanic activity on Io's surface, ejecting vast quantities of SO$_{2}$ into Io's tenuous atmosphere. This SO$_{2}$ is ionized into species such as S$^{2+}$ and subsequently captured by Jupiter's magnetosphere (Saur et al. 2004). The Io plasma torus is a relatively cool, low density plasma. Due to the many {\em in situ} density measurements obtained by various flyby missions, the Io plasma torus is an ideal object to study with our new set of collision strength data.

All line ratio calculations presented in this section have been undertaken with {\sc cloudy} C13 assuming a Maxwellian electron energy distribution for the Io torus. Although the Io torus is known to possess a suprathermal tail (see Feldman et al. 2004 and references therein), for the purpose of  
the present work a Maxwellian distribution should give a reasonable approximation, as we are primarily interested in comparisons with previous line ratio calculations which also adopted a Maxwellian distribution.

\subsubsection{FUSE observations in the 905--1187 {\AA} region}

In January 2001, an extensive examination of the west ansa of the Io plasma torus was undertaken
by FUSE over the wavelength range 905--1187\,\AA.  Spectra were obtained over five contiguous orbits and provided a noticeable improvement, with regard to signal-to-noise ratio, over previous observations. The spectra were studied in detail by Feldman et al. (2004), who used two S\,{\sc iii} n$_{e}$--sensitive line ratios to determine the electron density within the Io plasma torus, but found very large differences 
between the two results. However, since 2004, the atomic data for S\,{\sc iii} have undergone substantial improvements, and we hence re-examine the density diagnostic ratios presented by Feldman et al. (2004) using our new results.

\begin{figure}[h]
\centering
\includegraphics[width=\hsize]{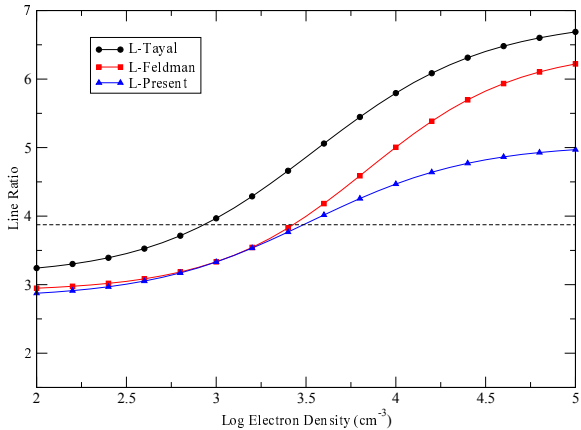}
\caption{The S\,{\sc iii} (1021.1 + 1021.3\,\AA)/1012\,\AA\ line intensity ratio plotted as a function of logarithmic electron density at an electron temperature of 50,000\,K. The dashed line represents the measured value from the 
FUSE spectrum of Feldman et al. (2004). The blue line with triangles is the L-Present calculations, the 
black line with circles is the L-Tayal values, and red line with squares the L-Feldman results.}
\label{felra1}
\end{figure}

One of the allowed line ratios reported by Feldman et al. (2004) is S\,{\sc iii} $\lambda\lambda$1021/$\lambda$1012, for which a measured ratio of 3.9 was determined from the FUSE spectrum. In Figure \ref{felra1} we show L-Present, L-Tayal and L-Feldman calculations for this ratio, where it can be seen that the present results are much smaller than the L-Tayal values, hence implying a greater electron density for the same measured ratio.
In the case of the Io plasma torus, L-Present yields an electron density of 10$^{3.5}$\,cm$^{-3}$, compared to a much smaller
value of 10$^{2.9}$\,cm$^{-3}$ from L-Tayal. 
With each flyby, a slightly different estimate 
was obtained for the electron density of the Io plasma torus, with values ranging from 10$^{3}$--10$^{3.6}$\,cm$^{-3}$ (Saur et al. 2004). The L-Present density lies within the experimental range, but that from L-Tayal is somewhat 
outside it. 

\begin{figure}[h]
\centering
\includegraphics[width=\hsize]{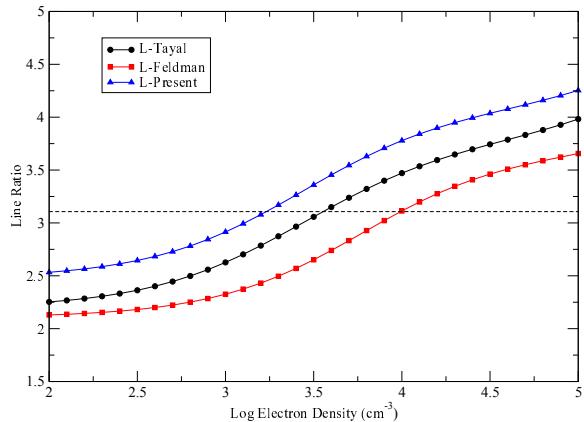}
\caption{Same as Figure 7 except for the S\,{\sc iii} 1077\,{\AA}/1012\,{\AA} line intensity ratio.}
\label{felra2}
\end{figure}

The second (weakly) density sensitive line ratio identified by Feldman et al. (2004) involves 
the S\,{\sc iii} 1077 and 1012 {\AA} lines. We present theoretical results in Figure \ref{felra2},
where we see that the ratio 
only varies by a factor of 2 between densities of 10$^{2}$ to 10$^{5}$\,cm$^{-3}$. Hence 
small changes in the atomic data and resultant theoretical ratios will lead to large variations in the derived density.
This was the problem encountered by Feldman et al. 
and is the reason that the L-Feldman results yield an unusually high density of 10$^{4}$\,cm$^{-3}$ for the measured ratio of 3.1. 
The L-Tayal ratios are smaller than the L-Present values and hence will yield a larger 
electron density. For the Io plasma torus, the L-Present data indicate n$_{e}$ = 10$^{3.2}$\,cm$^{-3}$, while the 
L-Tayal results imply n$_{e}$ = 10$^{3.6}$\,cm$^{-3}$. Both results lie within the density range determined from {\em in situ} measurements. However, the densities derived from the two line ratios using the L-Present calculations
(10$^{3.5}$ and 10$^{3.2}$\,cm$^{-3}$) are in much better agreement than those from L-Tayal (10$^{2.9}$ and 10$^{3.6}$\,cm$^{-3}$), highlighting the improvement in internal consistency provided by the current theoretical results.

The average electron density derived from the L-Present calculations is n$_{e}$ = 10$^{3.4}$\,cm$^{-3}$, in 
excellent agreement with the Voyager 1 {\em in situ} measurement of n$_{e}$ = 10$^{3.3}$\,cm$^{-3}$. 
This value will now be employed to generate 
a set of theoretical line intensity ratios for comparison with the observed results of Feldman et al. (2004). These
are presented in Table \ref{intensityfeldman}. Note that the
three S\,{\sc iii} lines from 1015.4 -- 1015.8\,{\AA} are blended in the FUSE spectra, and hence in Table 4 we compare theory and observation for the blended feature.
Similarly, the 1021.1 and 1021.3\,{\AA} lines are blended, as are 1126.5 and 1126.9\,{\AA}. 
It can be seen from Table \ref{intensityfeldman} that generally the present theoretical results lie 
closer to the observed values than those from CHIANTI. The average discrepancy between theory and observation for the current calculations is 12\%, as opposed to 23\%\ for the CHIANTI ratios. In particular, for the 
lines at 1126\,{\AA} we find a significant improvement over CHIANTI, reducing the discrepancy between theory and
observation from 30\% to 12\%. 

\begin{deluxetable}{l c c c c}
\tablewidth{0 pt}
\tablecaption{Theoretical and observed intensities for S {\sc iii} emission lines in FUSE spectra of the Io plasma torus.\tablenotemark{a}\label{intensityfeldman}}
\tablehead{
\colhead{Wavelength ({\AA})} & \colhead{Transition} & \colhead{Present\tablenotemark{b}} & \colhead{CHIANTI\tablenotemark{b}} & \colhead{Feldman\tablenotemark{c}}}
\startdata
1012.504 & 3s$^{2}$3p$^{2}$ $^{3}$P$_{0}$ -- 3s3p$^{3}$ $^{3}$P$_{1}$ & 0.31 & 0.34 & 0.32
\\
1015.487 + & 3s$^{2}$3p$^{2}$ $^{3}$P$_{1}$ -- 3s3p$^{3}$ $^{3}$P$_{0}$ + & 0.89 & 0.95 & 0.76
\\
1015.573 + & 3s$^{2}$3p$^{2}$ $^{3}$P$_{1}$ -- 3s3p$^{3}$ $^{3}$P$_{1}$ + 
\\
1015.783 & 3s$^{2}$3p$^{2}$ $^{3}$P$_{1}$ -- 3s3p$^{3}$ $^{3}$P$_{2}$ 
\\
1021.112 + & 3s$^{2}$3p$^{2}$ $^{3}$P$_{2}$ -- 3s3p$^{3}$ $^{3}$P$_{1}$ + & 1.46 & 1.57 & 1.25
\\
1021.328 & 3s$^{2}$3p$^{2}$ $^{3}$P$_{2}$ -- 3s3p$^{3}$ $^{3}$P$_{2}$ 
\\
1077.145 & 3s$^{2}$3p$^{2}$ $^{1}$D$_{2}$ -- 3s$^{2}$3p3d $^{1}$D$_{2}$ &  1.00 & 1.00 & 1.00 
\\
1121.760 & 3s3p$^{3}$ $^{3}$D$_{2}$ -- 3s$^{2}$3p4p $^{3}$P$_{2}$ & 0.004 &  0.004 & 0.004 
\\
1122.413 & 3s3p$^{3}$ $^{3}$D$_{3}$ -- 3s$^{2}$3p4p $^{3}$P$_{2}$ & 0.018 &  0.021 & 0.018 
\\
1126.536 + & 3s3p$^{3}$ $^{3}$D$_{1}$ -- 3s$^{2}$3p4p $^{3}$P$_{1}$ + & 0.029 &  0.023 & 0.033
\\
1126.879 & 3s3p$^{3}$ $^{3}$D$_{2}$ -- 3s$^{2}$3p4p $^{3}$P$_{1}$
\\
1128.497 & 3s3p$^{3}$ $^{3}$D$_{1}$ -- 3s$^{2}$3p4p $^{3}$P$_{0}$ & 0.022 &  0.014 &  0.035
\\
\enddata
\tablenotetext{a}{Intensities listed relative to I(1077.145\,{\AA}).}
\tablenotetext{b}{Theoretical results calculated at T$_{e}$ = 50,000\,K and n$_{e}$ = 10$^{3.4}$\,cm$^{-3}$.}
\tablenotetext{c}{Measured values from Feldman et al. (2004).}
\end{deluxetable}

\subsubsection{EUVE observations in the 370--735 {\AA} region}

The EUVE satellite contained three spectrographs which provided coverage of the 70--750\,{\AA} region. It 
observed the Io plasma torus over several orbits during a two-day period in 1993, and 
Hall et al. (1994b) studied the resulting emission line spectrum and reported numerous S\,{\sc iii} features.

The first region of interest in the Io spectrum was 480--489\,{\AA} (at shorter wavelengths no S\,{\sc iii} 
emission features were detected), where the 3s$^{2}$3p$^{2}$ $^{3}$P -- 3s3p4d $^{3}$P, $^{3}$D transitions 
give rise to many lines. As noted earlier (see Figure 5), very large differences (of up to a factor of 12)
were discovered between the present effective collision strengths and the results of Tayal \& Gupta (1999) for transitions to levels involving the 4d orbital. These discrepancies should in turn have a significant effect 
on the resulting line intensity ratios, as confirmed in Table 5 where we list both the theoretical and measured 
values. It can be seen from the table that for all of the predicted line intensity ratios, apart from those
involving 484.2, 485.8 and 486.1\,{\AA}, there are very large differences between the present calculations
and those from CHIANTI, of up to a factor of 3.

\begin{deluxetable}{l c c c c c c c }
\tablewidth{0 pt}
\tablecaption{Theoretical line intensity ratios for S\,{\sc iii} transitions between the 3s$^{2}$3p$^{2}$ and  3s$^{2}$3p4d levels.\tablenotemark{a}\label{intensityhall}}
\tablehead{
 & & \multicolumn{3}{c}{Present work\tablenotemark{b}} & \multicolumn{3}{c}{CHIANTI\tablenotemark{b}}
 \\
\colhead{Wavelength ({\AA})} & \colhead{Transition} & \colhead{$10^{3}$} & \colhead{$10^{3.5}$} & \colhead{$10^{4}$} & \colhead{$10^{3}$} & \colhead{$10^{3.5}$} & \colhead{$10^{4}$}}
\startdata
480.533 & $^{3}$P$_{0}$ - $^{3}$P$^{\mathrm{o}}_{1}$ & 0.167 & 0.137 & 0.120 & 0.524 & 0.329 & 0.254 \\
480.968 & $^{3}$P$_{1}$ - $^{3}$P$^{\mathrm{o}}_{0}$ & 0.311 & 0.398 & 0.420 & 0.944 & 1.000& 0.972 \\
481.234 & $^{3}$P$_{1}$ - $^{3}$P$^{\mathrm{o}}_{1}$  & 0.513 & 0.417 & 0.369 & 1.607 & 1.006 & 0.779 \\
482.565 & $^{3}$P$_{2}$ - $^{3}$P$^{\mathrm{o}}_{1}$ & 0.528 & 0.430 & 0.380 & 1.656 & 1.040 & 0.801 \\
484.172 &  $^{3}$P$_{0}$ - $^{3}$D$^{\mathrm{o}}_{1}$ & 1.237 & 0.778 & 0.559 & 1.230 & 0.607 & 0.392 \\ 
484.564 & $^{3}$P$_{1}$ - $^{3}$D$^{\mathrm{o}}_{2}$ & 1.000 & 1.000 & 1.000 & 1.000 & 1.000 & 1.000\\
484.874 & $^{3}$P$_{1}$ - $^{3}$D$^{\mathrm{o}}_{1}$ & 0.639 & 0.402 & 0.289 & 0.639 & 0.312 & 0.204 \\
485.255 & $^{3}$P$_{2}$ - $^{3}$D$^{\mathrm{o}}_{3}$ & 0.687 & 0.718 & 0.907 & 0.377 & 0.416 & 0.635 \\
485.818 & $^{3}$P$_{2}$ - $^{3}$D$^{\mathrm{o}}_{2}$  & 0.143 & 0.143 & 0.143 & 0.139 & 0.145 & 0.144 \\
486.140 & $^{3}$P$_{2}$ - $^{3}$D$^{\mathrm{o}}_{1}$ & 0.015 & 0.010 & 0.008 & 0.016 & 0.006 & 0.006 \\
\enddata
\tablenotetext{a}{Intensities listed relative to I(484.56\,{\AA}).}
\tablenotetext{b}{Theoretical results calculated at T$_{e}$ = 50,000\,K and at 3 electron densities of 10$^{3}$, 10$^{3.5}$ and 10$^{4}$\,cm$^{-3}$.}
 \end{deluxetable}

 Due to the large degree of blending in this spectral region it may be more appropriate to consider multiplet intensities. The present results indicate that the $^{3}$P -- $^{3}$D$^{\mathrm{o}}$ multiplet should be about a factor of 2.2 stronger than $^{3}$P -- $^{3}$P$^{\mathrm{o}}$. However, CHIANTI suggests that $^{3}$P -- $^{3}$D$^{\mathrm{o}}$ is actually about 30\%\ weaker than $^{3}$P -- $^{3}$P$^{\mathrm{o}}$. Unfortunately, the 
 lines are too blended to draw any realistic conclusions from the observational data reported by Hall et al. (1994b).
 
The next two EUVE regions of interest for S\,{\sc iii} transitions were 671--685\,{\AA} and 724--731\,{\AA}, and 
theoretical line intensity ratios for both spectral regions are presented in Table \ref{intensityhall2}. From the table
we can see that there is much better agreement between the present calculations and those from CHIANTI 
for these two spectral regions than was the case for the 480--489\,\AA\ wavelength range. 
The largest differences occur for transitions to the 3s$^{2}$3p4s levels, where for example the current 681.5/678.5 and 683.5/678.5 line ratios are typically 60\%\ larger than those from CHIANTI.

\begin{deluxetable}{l c c c c c c c c}
\tablewidth{0 pt}
\tablecaption{Theoretical line intensity ratios for S\,{\sc iii} transitions in the 671--685 {\AA} and 724--731 {\AA} spectral regions.\tablenotemark{a}\label{intensityhall2}}
\tablehead{
 & & \multicolumn{3}{c}{Present work\tablenotemark{b}} & \multicolumn{3}{c}{CHIANTI\tablenotemark{b}}
 \\
\colhead{Wavelength ({\AA})} & \colhead{Transition} & \colhead{$10^{3}$} & \colhead{$10^{3.5}$} & \colhead{$10^{4}$} & \colhead{$10^{3}$} & \colhead{$10^{3.5}$} & \colhead{$10^{4}$}}
\startdata
677.729 & 3s$^{2}$3p$^{2}$ $^{3}$P$_{0}$ - 3s$^{2}$3p3d $^{3}$D$^{\mathrm{o}}_{1}$ &  1.726 & 0.925 & 0.603 & 1.572 & 0.823 & 0.541\\
678.456 & 3s$^{2}$3p$^{2}$ $^{3}$P$_{1}$ - 3s$^{2}$3p3d $^{3}$D$^{\mathrm{o}}_{2}$ & 1.000 & 1.000 & 1.000 & 1.000 & 1.000 & 1.000 \\
679.104 & 3s$^{2}$3p$^{2}$ $^{3}$P$_{1}$ - 3s$^{2}$3p3d $^{3}$D$^{\mathrm{o}}_{1}$ & 1.615 & 0.865 & 0.564 & 1.471 & 0.770 & 0.507 \\
680.677 & 3s$^{2}$3p$^{2}$ $^{3}$P$_{2}$ - 3s$^{2}$3p3d $^{3}$D$^{\mathrm{o}}_{3}$ & 0.695 & 0.739 & 1.021 &  0.651 & 0.750 & 1.121 \\
680.925 & 3s$^{2}$3p$^{2}$ $^{3}$P$_{2}$ - 3s$^{2}$3p3d $^{3}$D$^{\mathrm{o}}_{2}$ & 0.478 & 0.478 & 0.478 & 0.478 & 0.478 & 0.478 \\
680.974 & 3s$^{2}$3p$^{2}$ $^{3}$P$_{1}$ - 3s$^{2}$3p4s $^{3}$P$^{\mathrm{o}}_{2}$ & 0.596 & 0.555 & 0.591 &  0.372 & 0.356 & 0.406 \\
681.489 & 3s$^{2}$3p$^{2}$ $^{3}$P$_{0}$ - 3s$^{2}$3p4s $^{3}$P$^{\mathrm{o}}_{1}$ & 0.817 & 0.443 & 0.317 & 0.468 & 0.270 & 0.211 \\
681.577 & 3s$^{2}$3p$^{2}$ $^{3}$P$_{2}$ - 3s$^{2}$3p3d $^{3}$D$^{\mathrm{o}}_{1}$ & 0.158 & 0.084 & 0.055 & 0.144 & 0.075 & 0.049 \\
682.879 & 3s$^{2}$3p$^{2}$ $^{3}$P$_{1}$ - 3s$^{2}$3p4s $^{3}$P$^{\mathrm{o}}_{1}$ & 0.019 & 0.010 & 0.007 & 0.011 & 0.006 & 0.005 \\
683.066 & 3s$^{2}$3p$^{2}$ $^{3}$P$_{1}$ - 3s$^{2}$3p4s $^{3}$P$^{\mathrm{o}}_{0}$ & 0.206 & 0.209 & 0.200 & 0.135 & 0.132 & 0.127 \\
683.461 & 3s$^{2}$3p$^{2}$ $^{3}$P$_{2}$ - 3s$^{2}$3p4s $^{3}$P$^{\mathrm{o}}_{2}$ & 0.024 & 0.022 & 0.024 & 0.015 & 0.014 & 0.016 \\
724.288 & 3s$^{2}$3p$^{2}$ $^{3}$P$_{0}$ - 3s3p$^{3}$ $^{3}$S$^{\mathrm{o}}_{1}$ & 0.321 & 0.235 & 0.218 & 0.278 & 0.203 & 0.193 \\
725.858 & 3s$^{2}$3p$^{2}$ $^{3}$P$_{1}$ - 3s3p$^{3}$ $^{3}$S$^{\mathrm{o}}_{1}$ & 0.901 & 0.660 & 0.612 & 0.779 & 0.570 & 0.541 \\
728.685 & 3s$^{2}$3p$^{2}$ $^{3}$P$_{2}$ - 3s3p$^{3}$ $^{3}$S$^{\mathrm{o}}_{1}$ & 1.326 & 0.972 & 0.901 & 1.147 & 0.839 & 0.796 \\
729.521 & 3s$^{2}$3p$^{2}$ $^{1}$D$_{2}$ - 3s$^{2}$3p4s $^{1}$P$^{\mathrm{o}}_{1}$ & 0.258 & 0.185 & 0.171 & 0.151 & 0.111 & 0.107 \\
\enddata
\tablenotetext{a}{Intensities listed relative to I(678.456\,{\AA}).}
\tablenotetext{b}{Theoretical results calculated at T$_{e}$ = 50,000\,K and at 3 electron densities of 10$^{3}$, 10$^{3.5}$ and 10$^{4}$\,cm$^{-3}$.}
\end{deluxetable}

Hall et al. (1994b) measured the following S\,{\sc iii} line ratios in the EUVE data:
   
 \begin{equation}
 R_{1}=\frac{3\mathrm{p}^{3} \  {}^{3}\mathrm{P} - 3\mathrm{d} \ {}^{3}\mathrm{D}^{\mathrm{o}}}{3\mathrm{p}^{3} \ {}^{3}\mathrm{P} - 4\mathrm{s} \ {}^{3}\mathrm{P}^{\mathrm{o}}} 
 \end{equation}
 
 \begin{equation}
  R_{2}=\frac{3\mathrm{p}^{3} \ {}^{3}\mathrm{P} - 3\mathrm{p}^{3} \ {}^{3}\mathrm{S}^{\mathrm{o}}}{3\mathrm{p}^{3} \ {}^{1}\mathrm{D} - 4\mathrm{s} \ {}^{1}\mathrm{P}^{\mathrm{o}} }
   \end{equation}
 
 and found $R_{1}$=3.8 and $R_{2}$=9.0. Our theoretical ratio for $R_{1}$ = 3.3, in good agreement with Hall et al. (1994b),
 while CHIANTI predicts a much larger value of $R_1$ = 4.9. In the case of $R_2$, we predict $R_2$ = 10.1 compared to $R_2$ = 14.5 from CHIANTI. Hence, for both ratios the present theoretical results are in better agreement with observations than are those from CHIANTI, which employ the effective collision strengths of Tayal \& Gupta (1999).
 
 \section{Conclusions}
 
Collision strengths have been generated for 1378 transitions in S\,{\sc iii}, with a total of 
53 fine-structure levels, arising from the 3s$^{2}$3p$^{2}$, 3s3p$^{3}$, 3s$^{2}$3p3d, 3s$^{2}$3p4s, 3s$^{2}$3p4p and 3s$^{2}$3p4d configurations, included in the calculation. These atomic data were produced using the {\sc rmatrxii}, {\sc fine} and {\sc pstgf} codes, and subsequently averaged over a Maxwellian distribution of electron energies to obtain the corresponding effective collision strengths. 
 
 The CHIANTI database currently adopts the collision strength data of Tayal \& Gupta (1999). However, a comparison of the present results with those of Tayal \& Gupta has highlighted some large differences between the two calculations.
 In particular, there are major discrepancies 
 for transitions between the 3s$^{2}$3p$^{2}$ and 3s$^{2}$3p4d levels, which are important as 
 they have been observed in spectra of the Io plasma torus. Noticeable differences are also found for transitions from the 3s$^{2}$3p$^{2}$ levels to the 3s$^{2}$3p4s levels. However, there is excellent agreement for many of the lower lying-transitions such as 3s$^{2}$3p$^{2}$ -- 3s3p$^{3}$.
 
   The present set of effective collision strengths and the A-value data from the CHIANTI Version 7.1 database were incorporated into CLOUDY C13. Theoretical emission line intensity ratios were produced for a number of transitions 
  observed in spectra of the Io plasma torus. Initially, FUSE spectra were examined and the observational data
  of Feldman et al. (2004) were compared with the theoretical values. The line intensity ratios 
  predicted by the present atomic data were found to show an average discrepancy of 12\%\ with the observed FUSE ratios, compared to an average difference between theory and observation of 23\%\ for the CHIANTI predictions.
Secondly, comparisons were made for 
two line ratios obtained by Hall et al. (1994b) from EUVE spectra of the Io plasma torus. Again, it was found that the present theoretical results were in better agreement with Hall et al. (1994b) than those obtained with the effective collision strength data of Tayal \& Gupta (1999). Hence, the present set of collision strengths are recommended as the most reliable for use in astrophysical modelling. Consequently, they will  be included in a future release of the CHIANTI database.

\acknowledgements
 
 This work has been supported by PPARC/STFC under the auspices of a rolling grant, and by the Leverhulme Trust. M.F.R. Grieve is funded by a DEL studentship.
The authors wish to acknowledge  G.J. Ferland and M. Lykins of the University of Kentucky (USA) for extending CLOUDY to facilitate direct inclusion of our atomic data.  CHIANTI is a collaborative project involving George Mason University, the University of Michigan (USA) and the University of Cambridge (UK).



\begin{thebibliography}{99}
\bibitem[Ballance(2004)]{pstgf}
Ballance, C. P., \& Griffin, D. C. 2004, J. Phys. B, 37, 2943
\bibitem[Binette(2012)]{binette} 
Binette, L., et al. 2012, A\&A, 547, A29
\bibitem[Burgess(1992)]{BurTul}
Burgess, A., \& Tully J.A. 1992, Astron. \& Astrophys., 254, 436.
\bibitem[Burke(1994)]{burke}
Burke, P. G., Burke, V. M., \& Dunseath, K. M. 1994, J. Phys. B: At. Mol. Opt. Phys., 27, 5341
\bibitem[Broadfoot(1979)]{broadfoot}
Broadfoot, A. L., et al. 1979, Science, 204, 979
\bibitem[Curdt(2004)]{curdt}
Curdt, W., Landi, E., \& Feldman, U. 2004, A\&A, 427, 1045
\bibitem[Dere(1997)]{Dere97}
Dere, K. P., Landi E., Mason, H. E., Monsignori Fossi, B. C., \& Young, P. R. 1997, Astron. Astrophys. Suppl. Ser., 125, 149.
\bibitem[Dong(2011)]{DongDrain}
Dong, R., \& Draine, B. T. 2011, \apj, 727, 35.
\bibitem[Fang(2011)]{fang}
Fang, X., \& Liu, X.-W. 2011, MNRAS, 415, 181
\bibitem[Feldman(2004)]{Feld04}
Feldman, P. D., Strobel, D. F., Moos, H. W., \& Weaver, H. A. 2004, \apj, 601, 583.
\bibitem[Ferland(2013)]{ferland}
Ferland, G. J., et al. 2013, RevMexAA, in press (arXiv:1302.4485)
\bibitem[Fischer(2006)]{fischer} 
Froese Fischer, C., Tachiev, G., \& Irimia, A. 2006, ADNDT, 92, 607
\bibitem[Galavis(1995)]{galavis}
Galav\'{i}s, M. E., Mendoza, C., \& Zeippen, C. J.  1995, Astron. Astrophys. Suppl. Ser., 111, 347
\bibitem[Hall(1994a)]{Hall94a}
Hall, D. T., Bednar, C. J., Durrance, S. T., Feldman, P. D., \& McGrath, M. A. 1994a, \apj, 420, L45.
\bibitem[Hall(1994b)]{Hall94b}
Hall, D. T., Gladstone, G. R., Moos, H. W., Bagenal, F., Clarke, J. T., Feldman, P. D., McGrath, M. A., Schneider, N. M., Shemansky, D. E., Strobel, D. F., \& Waite, J. H. 1994b, \apj, 426, L51.
\bibitem[Hibbert(1975)]{hibbert}
Hibbert, A. 1975, Comput. Phys. Comm., 9, 141
\bibitem[Huang(1985)]{huang}
Huang, H.-N., 1985, ADNDT, 32, 503
\bibitem[Hudson(2012)]{hudson12}
Hudson, C. E., Ramsbottom, C. A., \& Scott, M. P. 2012, \apj, 750, 65
\bibitem[Landi(2012)]{Landi12}
Landi, E., Del Zanna, G., Young, P. R., Dere, K. P., \& Mason, H.E. 2012, \apj, 744, 99.
\bibitem[NIST(2013)]{nist}
NIST Atomic Spectra Database (version 3.1.5), [Online] Available: {\tt http://physics.nist.gov/asd3} [2010, March 29]. Kramida, A., Ralchenko, Yu., Reader, J., and NIST ASD Team (2012). National Institute of Standards and Technology, Gaithersburg, MD.
\bibitem[Otsuka(2010)]{Otsuka}
Otsuka, M., Tajitsu, A., Hyung, S., \& Izumiura, H. 2010, \apj, 723, 658.
\bibitem[Saur(2004)]{saur}
Saur, J., Neubauer, F. M., Connerney, J. E. P., Zarka, P., \& Kivelson, M. G. 2004, Jupiter. The planet, satellites and magnetosphere, Vol. 1 (1st ed.; Cambridge, UK: Cambridge University Press)
\bibitem[simpson(1997)]{simpson}
Simpson, J. P., et al. 1997, ApJ, 487, 689
\bibitem[Tayal(1997)]{tay} 
Tayal, S. S., 1997, ADNDT, 67, 331
\bibitem[Tayal(1999)]{tayal}
Tayal, S. S., \& Gupta, G. P.  1999, \apj, 526, 544 
\end{thebibliography}
\end{document}